\documentclass[a4paper]{article}

\usepackage{eqis}

 \newcommand{\hs}[1]{\hspace*{ #1 mm}}
 \newcommand{\vs}[1]{\vspace*{ #1 mm}}

 \newcommand{\nat}{\mathbb{N}}

 \newcommand{\ie}{\textrm{i.e.},\hspace*{2mm}}

 \newcommand{\p}{\mathrm{P}}
 \newcommand{\np}{\mathrm{NP}}

 \newcommand{\pp}{\mathrm{PP}}

 \newcommand{\bqp}{\mathrm{BQP}}

 \newcommand{\fp}{\mathrm{FP}}

 \newcommand{\fbqp}{\mathrm{FBQP}}

 \newcommand{\cd}{\mathrm{CD}}

 \newcommand{\ic}{\mathrm{IC}}
 \newcommand{\kc}{\mathrm{C}}

 \newcommand{\qc}[2]{\mathrm{QC}_{ #1 }^{ #2 }}
 \newcommand{\qqc}[2]{\mathrm{qQC}_{ #1 }^{ #2 }}
 \newcommand{\dc}[2]{\mathrm{DC}_{ #1 }^{ #2 }}
 
 \newcommand{\ceilings}[1]{\lceil #1 \rceil}

 \newcommand{\qubit}[1]{| #1 \rangle}

\newcommand{\ignore}[1]{}

\begin{document}

\title{
   Quantum Minimal One Way Information: \\
Relative Hardness and Quantum Advantage of Combinatorial Tasks
}

\author{
  Harumichi Nishimura 
  \affiliation{1}
  \email{hnishimura@qci.jst.go.jp}
  \and 
  Tomoyuki Yamakami
  \affiliation{2}
  \email{TomoyukiYamakami@TrentU.CA}
}

\address{1}{
  ERATO Quantum Computation and Information Project\\
  Japan Science and Technology Agency, Kyoto, 602-0873 Japan
}

\address{2}{
  Computer Science Program, Trent University\\
  Peterborough, Ontario, Canada K9J 7B8
}

\abstract{
Two-party one-way quantum communication has been extensively studied in the recent literature. 
We target the size of minimal information that is necessary for a feasible party  
to finish a given combinatorial task, such as distinction of instances, using one-way communication from another party. 
This type of complexity measure has been studied under various names: advice complexity, 
Kolmogorov complexity, distinguishing complexity, and instance complexity. 
We present a general framework focusing on underlying combinatorial takes to study these complexity measures using quantum information processing. 
We introduce the key notions of relative hardness and quantum advantage, 
which provide the foundations for task-based quantum minimal one-way information complexity theory. 
}


\maketitle

Many branches of complexity theory can be explained under a broader concept of ``communication'' among multiple parties. 
Of all possible communication patterns, we restrict ourselves within two-party one-way communication 
because of its simplicity. Such a communication model has recently attracted much attention to the quantum setting of one-way communication complexity, 
Kolmogorov complexity, and advice complexity. 
We attempt to develop a coherent theory of the {\em size complexity} of minimal one-way information 
that is necessary to complete a given {\lq\lq}task{\rq\rq} by a feasible quantum algorithm. For simplicity, we set our alphabet $\Sigma$ to be $\{0,1\}$ and denote the empty string by $\lambda$.
Of all possible tasks, our attention is focused on the following types of tasks, called {\em combinatorial tasks}, which are multi-valued total functions $F$ from $\Sigma^*\times\Sigma^*$ 
to $\Sigma^*\cup\{\bot\}$, where $\bot$ is a special value meaning both ``I don't know'' and ``it doesn't halt.''  For readability, we often abbreviate $F(x,z)$ as $F[x](z)$. The {\em size} of instance $(x,z)$ of the task $F$, which is a natural number, is succinctly denoted $s_{F}[x](z)$ and called the {\em size function} of $F$. 
For convenience, we assume that $s_F[x](z)\geq |z|$ for all strings $z$. 
Typical examples of combinatorial tasks are {\lq\lq}generation{\rq\rq} $Gen$, {\lq\lq}distinction{\rq\rq} $Dist$, and {\lq\lq}evaluation{\rq\rq} $Eval$.
The task of generation $Gen$ is defined as $Gen[x](\lambda)=\{x\}$ 
and $Gen[x](z)=\Sigma^*\cup\{\bot\}$ for every nonempty string $z$. 
The size function of $Gen[x]$ is given as $s_{Gen}[x](z)=|x|+|z|$ for all strings $x$ and $z$. 
The task of distinction $Dist$ of $x$ is defined as $Dist[x](x)=\{1\}$ and $Dist[x](z)=\{0\}$ for all $z\in\Sigma^*\setminus\{x\}$. 
Its size function is given as $s_{Dist}[x](z)=|z|$ for all strings $x$ and $z$. 
For any given function $f$ from $\Sigma^*$ to $\Sigma^*$, the task of evaluation of $f$ is defined as  
$Eval_{f}[x](x)=\{f(x)\}$ and $Eval_{f}[x](z)=\{f(z),\bot\}$ for all $z\in\Sigma^*$ 
with the size function $s_{Eval_{f}}[x](z)=|z|$ for any string $z$. 

Now, we introduce the notion of task-based minimal one-way information complexity that provides a general framework for a study of a one-way communication model. 
We begin with the deterministic version of minimal one-way information complexity. 
First, we fix a universal deterministic Turing machine $U_c$. 
Let $x,y$ be any strings, $t$ be any function in $\nat^{\nat}$, and $F$ be any combinatorial task 
with its size function $s_F$. 
The {\em $t$-time deterministic minimal one-way information complexity of $F$ on $x$ conditional to $y$}, 
succinctly denoted $\dc{F}{t}(x|y)$, is the minimal length of a binary string $p$ 
such that, on input $(p,z,y)$, $U_c$ outputs an element of the set $F[x](z)$ 
within $t(s_{F}[x](z)+|y|)$ steps for any string $z\in\Sigma^*$. 
Whenever $y$ is the empty string, we omit $y$ and write $\dc{F}{t}(x)$. 
By taking the aforementioned combinatorial tasks, we immediately obtain the following existing complexity notions (see \cite{LV97}): 
time-bounded Kolmogorov complexity $\kc^{t}(x|y) = \dc{Gen}{t}(x|y)$, 
distinguishing complexity $\cd^{t}(x|y) = \dc{Dist}{t}(x|y)$,  and 
instance complexity $\ic^{t}(x:A) = \dc{Eval_{c_A}}{t}(x)$. 

Naturally induced from $\dc{F}{t}(x|y)$ by simply replacing $U_c$ with a fixed universal quantum Turing machine $U$, 
we can introduce the quantum analogue $\qc{F,\epsilon}{t}(x|y)$, where $\epsilon$ is an upper bound of the error probability that $U$ fails to complete the task. Moreover, we introduce the notion $\qqc{F,\epsilon}{t}(x|y)$ by allowing the use of a {\em qustring} (\ie a quantum state) $\qubit{\phi}$  
instead of a classical string $p$. Whenever $\epsilon=1/3$, we suppress subscript $\epsilon$. Note that the time-unbounded version $\qqc{Gen,\epsilon}{\infty}(x|y)$ corresponds to quantum Kolmogorov complexity with bounded fidelity 
of Berthiaume, van Dam, and Laplante  
except for the way to measure an error rate of generating classical strings. 
On the contrary, $\qc{Gen,\epsilon}{\infty}(x|y)$ corresponds to the quantum Kolmogorov complexity with classical information 
of Vit{\'a}nyi although, in his definition, an error rate is incorporated into his Kolmogorov complexity measure.  

The choice of an error bound $\epsilon$ in the above definition is not important for the case of classical information since we can reduce error probability at the cost of a constant additive term. 
In the quantum case, however, we might possibly need to pay the cost of a constant multiplicative term 
due to the {no-cloning theorem}. 

\begin{lemma}\label{change-error}{\rm [Amplification Lemma]}\hs{1}
Let $\epsilon,\epsilon'$ be any real numbers with $0<\epsilon'\leq \epsilon<1/2$, 
let $t$ be any function in $\nat^{\nat}$, and let $F$ be any combinatorial task. 
(1) There exists an absolute constant $c\geq0$ such that, for every pair of strings $x$ and $y$, 
$\qc{F,\epsilon'}{t'}(x|y)\leq \qc{F,\epsilon}{t}(x|y)+c$, where $t'(n)=d\cdot t(n)^2+d$ 
for a certain constant $d$ (depending on $\epsilon$ and $\epsilon'$). (2) There exists an absolute constant $c\geq0$ such that, 
for every pair of strings $x$ and $y$, $\qqc{F,\epsilon'}{t'}(x|y)\leq k\cdot \qqc{F,\epsilon}{t}(x|y)+c$, 
where $k=\ceilings{(2\log{\epsilon'}+2)/(\log\epsilon+\log(1-\epsilon)+2)}$ and  $t'(n)=d\cdot t(n)^2+d$ for a certain constant $d$.
\end{lemma}

Note that the bound in Lemma \ref{change-error}(2) is optimal for the task $Dist$ because, to reduce the error probability from a value $\epsilon'$ 
to another value $\epsilon$, we need $\Theta(\log\epsilon/\log\epsilon')$ copies of a minimal qustring 
to complete the task $Dist$ with error $\epsilon'$. 

We present lower bounds of the quantum minimal one-way information complexity of generation and distinction by quantum information.

\begin{lemma}
Let $g$ be any function from $\nat$ to $\nat\setminus\{0,1\}$, and $\epsilon$ be any real number in $[0,1/2)$. 
(1) There exists a constant $c\geq1$ such that, for any sufficiently large $n$, there are at least $2^n(1-2^{-\frac{g(n)-2}{1-\epsilon}}
+2^{-(\epsilon n+g(n)+c)}) -1$ strings $x$ of length $n$ satisfying $\qqc{Gen,\epsilon}{\infty}(x)\geq \ceilings{(1-\epsilon)n}-g(n)$. 
(2) For any $\delta>0$ and any sufficiently large $n$, at least $2^n(1-2^{-g(n)})$ strings $x$ of length $n$ satisfy 
$\qqc{Dist,\epsilon}{\infty}(x)\geq (1-\delta)\log(n-g(n))$. 
\end{lemma}

The quantum minimal one-way information complexity gives a unique way to look into the structure of each individual combinatorial task. 
With this complexity measure, we are to classify the combinatorial tasks by simply comparing among their complexity values. 
It is useful to introduce a simple binary relation that tells which of two given tasks has ``smaller'' complexity 
than the other. We introduce such a relation under the name of {\em relative hardness}. 
The relative hardness relation $\leq^{QC}$, its qustring version $\leq^{qQC}$ 
and their ``infinitely-often'' versions are defined between two combinatorial tasks $F$ and $F'$ as follows. 
We write $F'\leq^{QC} F$ (resp.\ $F'\leq^{QC}_{\mathrm{io}} F$) if, 
for any polynomial $t$, there exist a constant $c\geq0$ and a polynomial $t'$ 
such that $\qc{F'}{t'}(x)\leq \qc{F}{t}(x)+c$ for all but finitely many strings $x$ 
(resp.\ infinitely many strings $x$). 
We can similarly define the relations $F'\leq^{qQC} F$ and $F'\leq^{qQC}_{\mathrm{io}}F$ by replacing $\qc{}{}$ by $\qqc{}{}$.  
In addition, we define the relation $F'<^{QC} F$ if $F'\leq^{QC} F$ and $F\not\leq^{QC}F'$. 
Likewise, the relation $F'<^{qQC}F$ is defined.  

All of the above relations form {\em partial orderings}. 
For any combinatorial task $F$ in $\fbqp$ (the function class corresponding to $\bqp$) 
with size function $s_F[x](z)\geq |x|+|z|$, it follows that $F\leq^{QC}Gen$; in other words,
 generation is at least as hard as $F$ by classical information. 
Furthermore, generation seems to require more information than distinction. 
In fact, we can easily prove that $Dist\leq^{QC}Gen$ and $Dist\leq^{qQC}Gen$.  
Using the notions of quantum fingerprinting and the Holevo bound, we also obtain $Dist<^{qQC}Gen$. 
Although we do not know whether $Dist<^{QC}Gen$, we can construct an oracle relative to which $Dist$ is strictly easier than $Gen$. 
For any oracle $A$, the relation $\leq^{QC,A}$ denotes the relativized version of $\leq^{QC}$ relative to $A$.

\begin{theorem}\label{dis-less-gen}
There exists a recursive oracle $A$ such that $Dist<^{QC,A}Gen$.
\end{theorem}

Next, we discuss the relative hardness of $Eval$ compared to $Gen$. 
We write $\leq^{DC}_{\mathrm{io}}$ for the deterministic version of $\leq^{QC}_{\mathrm{io}}$. 
If $A$ satisfies $Gen\leq^{DC}_{\mathrm{io}} Eval_A$, then $A$ seems very difficult to compute. 
This intuition leads to define the set $HARD=\{A\in\mathrm{REC}\mid Gen\leq^{DC}_{\mathrm{io}} Eval_{A}\}$, 
where $\mathrm{REC}$ denotes the class of all recursive sets. 
It is, however, unknown whether any recursive set outside of $\p$ belongs to $HARD$. 
This turns out to be equivalent to the so-called {\em instance complexity conjecture} of Orponen, Ko, Sch{\"o}ning, and Watanabe. Fortnow and Kummer showed that the instance complexity conjecture holds if $\p=\np$. 
Similarly, we can raise the question of whether any recursive set outside of $\bqp$ belongs to 
the set $QHARD=\{A\in\mathrm{REC}\mid Gen\leq^{qQC}_{\mathrm{io}} Eval_A\}$. 
The following theorem implies that any recursive set outside of $\bqp$ belongs to $QHARD$ 
under the assumption that $\bqp=\pp$. 

\begin{theorem}\label{eval-less-gen}
If $\bqp=\pp$, then $\bqp=\{A\in\mathrm{REC}\mid Gen\not\leq^{qQC}_{\mathrm{io}} Eval_A\}$.
\end{theorem}

Relative hardness compares between two combinatorial tasks. 
It is also important to make a comparison between $\qc{F,0}{t}(x|y)$ and $\dc{F}{t'}(x|y)$ 
for the same combinatorial task $F$. 
What is the advantage of using quantum computation with classical (or quantum) information rather than 
deterministic computation with classical information? We formalize such an advantage under the term ``quantum advantage.''  
Let $k$ be any function in $\nat^{\nat}$ and let $F$ be any combinatorial task. 
We say that $F$ has {\em quantum $k(n)$-advantages over DC by classical information} 
if there exists a polynomial $t$ such that, for every polynomial $t'$, 
$k(\qc{F,0}{t}(x))\leq \dc{F}{t'}(x)$ for infinitely many strings $x$.  
Similarly, we define the term {\em quantum $k(n)$-advantages over DC by quantum information} 
by replacing the inequality $k(\qc{F,0}{t}(x))\leq \dc{F}{t'}(x)$ by $k(\qqc{F,0}{t}(x))\leq \dc{F}{t'}(x)$. 
We often drop the term ``over DC'' for simplicity.

Applying results on one-way quantum communication complexity, we can show, for instance, the existence of a task $F$ that has quantum $2^{n/d}$-advantages 
by quantum information for a certain constant $d>0$; however, no task 
is known to have quantum advantages by classical information.  
For a certain combinatorial task $F$, we can find a relativized world where $F$ truly 
possesses quantum advantages even by classical information.  

\begin{theorem}\label{qcg-advantage}
Let $F$ be any combinatorial task in $\fp$ that satisfies $Dist\leq^{DC}F$. 
Relative to a certain oracle, $F$ has quantum $2_n$-advantages by classical information, 
where $2_n$ is defined inductively by $2_0=1$ and $2_n=2^{2_{n-1}}$ for each positive integers $n$.
\end{theorem}

Is there any ``simple'' set $A$ that makes $Eval_A$ possess high quantum advantages? 
This question has a direct connection to the $\p=?\bqp$ question.
 
\begin{proposition}\label{eval-pbqp} 
The following three statements are equivalent. (1) $\p\neq\bqp$. 
(2) There exists a set $A\in\bqp$ such that $Eval_{A}$ has quantum $\omega(n)$-advantages 
by classical information. 
(3) There exists a set $A\in\bqp$ such that $Eval_{A}$ has quantum $\omega(n)$-advantages by quantum information.
\end{proposition}

\bibliographystyle{alpha}


\end{document}